\documentstyle[preprint,aps,epsf]{revtex}
\def\setepsfscale#1{\def\epsfsize##1##2{#1##1}}

\begin{document}
\draft
\preprint{\vbox{
\hbox{IFT-P.023/94}
\hbox{IFUSP/P-1117}
\hbox{DCR/TH 04-94}
\hbox{hep-ph/9407333}
\hbox{July 1994}
}}
\title{
Three generation vacuum oscillations and the solar neutrino problem }
\author{M.M. Guzzo}
\address{
 Instituto de F\'\i sica Gleb Wataghin \\
Universidade Estadual de Campinas, UNICAMP\\
13083-970 -- Campinas, SP\\
Brazil}
\author{ O.L.G. Peres, V. Pleitez }
\address{ Instituto de F\'\i sica Te\'orica\\
Universidade Estadual Paulista\\
Rua Pamplona, 145\\
01405-900 -- S\~ao Paulo, SP\\
Brazil}
\author{R. Zukanovich Funchal }
\address{ Instituto de F\'\i sica da Universidade de S\~ao Paulo\\
01498-970 C.P. 20516 -- S\~ao Paulo, SP\\
Brazil}
\maketitle
\newpage
\begin{abstract}
We investigate the solar neutrino problem in the
scenario of three generation neutrino oscillation hypothesis,
taking into account other phenomenological constraints to the neutrino mixing
and mass parameters.
\end{abstract}
\narrowtext
Although direct measurements of the neutrino masses are all consistent
with zero~\cite{numass}, it is well known that some issues, as the
solar~\cite{exp1} and atmospheric~\cite{exp2} neutrino problems, may be
an indication of nonvanishing neutrino masses once that an economical
and successful way to understand both results relies on the
neutrino oscillation hypothesis~\cite{pal}. Even neutrino oscillations in
vacuum continue to be quoted~\cite{varios} as a possible solution to
the solar neutrino problem~\cite{exp1}. Most of these analysis deal
with two neutrino oscillations, based on the simplified assumption
that only the mixing of the electron neutrino with another active or
sterile neutrino, as well as their squared mass differences are
nonvanishing parameters generating neutrino oscillations.

Nevertheless, there is no reason, in principle, for not considering
three generation neutrino oscillations in the interpretation of solar
neutrino data~\cite{three}. Moreover, some experimental evidence has
been accumulated for the existence of three light
neutrinos~\cite{pdg}.

The main difficulty concerning three generation neutrino oscillations
is connected with the appearance of too many free oscillating
parameters, namely, three mixing angles, one phase and two neutrino
squared mass differences which can not be phenomenologically fixed
taking into account only oscillation effects. In fact, when three
neutrino generation oscillation phenomenon is considered in
literature such problem is usually overcome fixing arbitrarily some
of the free parameters~\cite{three}.

Nevertheless, in Ref.~\cite{tau4}, assuming the minimal extension of
the standard electroweak model~\cite{ws}, when only three
right-handed neutrino singlets are introduced to generate Dirac
neutrino masses, two mixing angles and one neutrino mass were
constrained by experimental data from accelerators, reactors and
underground facilities. It is interesting to emphasize that these
angles and mass not include the values of these parameters which
would lead to the limit where three generation case behaves as a
usual two generation oscillating system. From this result we can
conclude that the three generation oscillations are a phenomenological
necessity.

In this paper we analyse the three generation neutrino oscillations as
a solution to the solar neutrino problem taking into account the
phenomenological constraints from Ref.~\cite{tau4} for some of the
mixing angles and neutrino masses entering the three generation
neutrino oscillation phenomenon.

Considering the very little restrictive hierarchy among neutrino
masses $m_1\alt m_2\ll m_3$, using the Maiani parametrization
{}~\cite{pdg,maiani} of the mixing matrix
\begin{equation}
V=\left(
\begin{array}{ccc}
c_\theta c_\beta \; \; \; & s_\theta c_\beta  \; \; & s_\beta \\
-s_\theta c_\gamma-c_\theta s_\gamma s_\beta \; \; \; & \; \; \;
c_\theta c_\gamma
-s_\theta s_\gamma s_\beta\; \; \;  & \; \; \;
s_\gamma c_\beta  \\
s_\theta s_\gamma -c_\theta c_\gamma s_\beta \; \; \; & \; \; \;
-c_\theta s_\gamma
-s_\theta c_\gamma s_\beta\; \; \;  & \; \; \;
c_\gamma c_\beta
\end{array}
\right)
\label{ckm}
\end{equation}
and investigating $\tau$ leptonic decay, pion decay, $Z^0$ invisible
width, $\tau$ decay end--point into five pions and assuming world
average data for the ratio $G_\tau/G_\mu$, the lower masses $m_1,m_2$
and one mixing angle $\theta$ remain undetermined, but $m_3\sim165$
MeV, $11.54^o<\beta<12.82^o$ and $\gamma<4.05^o$.
Thus, we have one mixing angle $\theta$ and the two lightest neutrino
squared mass difference $\delta m^2=m_2^2-m_1^2$ to be determined in
neutrino oscillation processes such as the solar neutrino issue.
Note that the phase which would appear in Eq.~(\ref{ckm}) is irrelevant for
the case of solar neutrino so we have ignored it.

The solar neutrino problem has been confirmed by many experiences.
In the following we will consider experimental data from
 Homestake, Kamiokande and $^{71}Ga$ experiences~\cite{exp1}. Each of
them are
sensitive to different types of neutrinos. In the Homestake experience
$78\%$ of the neutrinos detected are the so-called $^8B$ neutrinos,
$14\%$ are $^7Be$ ones and about $4\%$ of them  are $^{15}O$
neutrinos~\cite{bu88}.
Other sources of
solar neutrinos contribute significantly less to the total theoretical
capture rate in the $Cl$-detector
than the total uncertainties involved in the
calculations and will be neglected. This approximation is justified
since for the case of $Cl$ detector  the theoretical
uncertainties are about $33\%$. Kamionande is sensitive only to the $^8B$
neutrinos. And, finally, neutrinos detected by $^{71}Ga$ experiences
are composed by
$26\%$, $11\%$, $5\%$ and $54\%$ of $^7Be$, $^8B$, $^{15}O$ and $pp$
neutrinos, respectively. Again we are neglecting sources of neutrinos
which contribute to the total $Ga$
detector  rate significantly less than the total theoretical uncertainties
($15\%$).

Neutrinos produced in different reactions have different energies. While
$^7Be$ neutrinos are almost monochromatic~\cite{ba},
neutrinos produced in other source-reactions have different energy
spectra \cite{bu88} which have to be considered since, as we will see
in the following, the survival probability of the solar  neutrinos
is sensitive to their energy $E$ or their momentum~$p$.

We can compare the theoretical neutrino flux ($\phi_{th}$) calculated
from the solar standard model~\cite{bu88} with the observed flux
($\phi_{exp}$)
measured by each experiment~\cite{exp1}. The ratio
$R=\phi_{exp}/\phi_{th}$ is given by
$R(\mbox{Homestake})=0.28\pm0.04$,
$R(\mbox{Kamiokande})=0.49\pm 0.12$, and for the two experiences
based on $^{71}Ga$ detectors:
$R(\mbox{Gallex})=0.66\pm0.12$ and $R(\mbox{Sage})=0.58\pm0.11$~\cite{new}
(Note that these two last numbers are compatible and we will
consider in the present
analysis only the Gallex result and do not use the
corresponding weighted average result.)

Here we introduced the survival transition probability for the electron
neutrino observed at a point $x$ if neutrinos were
produced deep in the Sun at a point $x_0$ in reaction $X$:
\begin{equation}
P^J(X)=\sum_{E_i>E^J_{thre}}^Ef^X(E_i)P_{\nu_e\to\nu_e}(E_i,\delta
m^2,\theta,R).
\label{pes}
\end{equation}
$J=H,K$ and $G$ index indicates Homestake, Kamiokande and Gallex.
The threshold energy for each one of these experiences and the energy
spectrum of neutrinos produced in reaction $X$ are denoted  by
$E^J_{thre}$ and $f^X(E_i)$, respectively. The spectral functions
$f^X(E_i)$ are given in Ref.~\cite{bu88b}.

The probability of finding a neutrino $\nu_e$ after a
length $x-x_0$ if at the origin it was a $\nu_e$ is
$P_{\nu_e\to\nu_e}(E,\delta
m^2,\theta,x)=\vert\langle\nu_e(x)\vert\nu_e(x_0)\rangle\vert^2$,
or explicitly
\begin{equation}
P_{\nu_e\to\nu_e}(E,\delta
m^2,\theta,r)=
1-2c_\beta^2s_\beta^2\left(1-\cos\frac{2\pi r}{L}\right)
-2c_\theta^2s_\theta^2c_\beta^4\left(1-\cos\frac{2\pi r}{L_{12}}\right),
\label{ee}
\end{equation}
where $r\equiv x-x_0\approx L_\odot$, being $L_\odot$ the  Earth-Sun distance.
For the mass range we are considering here,
neutrino oscillations occur with essentially two wavelengths. We have
defined in Eq.~(\ref{ee}) $ L_{ij}=2\pi/(E_i-E_j)$
that is, $L_{12}=4\pi p/\delta m^2=2.5(p/\mbox{MeV})/(\delta
m^2/\mbox{eV}^2)$ meter, and
$L_{13}=L_{23}=L=2\pi/m_{3}=1.24/(m_{3}/\mbox{MeV}) \times10^{-12}$ meter.
The shorter wavelength is of order of $10^{-15}$ meter, for
$m_3\sim165$ MeV. With respect
to the solar neutrinos $r\sim10^{11}$ meter, with this condition we can average
out
the cosine term involving $L$ and Eq.~(\ref{ee}) becomes
\begin{equation}
P_{\nu_e\to\nu_e}(E,\delta
m^2,\theta,r)=
1-2c_\beta^2s_\beta^2
-2c_\theta^2s_\theta^2c_\beta^4\left(1-\cos\frac{2\pi r}{L_{12}}\right).
\label{ee2}
\end{equation}
We can also write down the transition probability for
each experience. For Homestake (H)
\begin{mathletters}
\label{pro}
\begin{equation}
R(\mbox{Homestake})=0.78P^H(^8B)+0.14P^H(^7Be)+0.04P^H(^{15}O).
\label{proho}
\end{equation}
The neutrino flux measured by Kamiokande
facilities is not merely the electron neutrino one
since detector electrons will interact with other neutrino flavors via neutral
currents. For energies involved in the solar neutrino experiences, the
$\nu_e$-electron scattering cross section is about seven times larger than
other neutrino flavor ($\nu_\mu$-electron and $\nu_\tau$-electron) cross
sections. Hence, for Kamiokande (K) we have
\begin{equation}
R(\mbox{Kamiokande})=P^K(^8B) + \frac{1}{7}[1-P^K(^8B)].
\label{proka}
\end{equation}
Finally, for $^{71}Ga$ detectors (G)
\begin{equation}
R(\mbox{G})=0.26P^G(^8B)+0.11P^G(^7Be)
+0.05P^G(^{15}O)+0.54P^G(pp).
\label{proga}
\end{equation}
\end{mathletters}

We can introduce Eq. (\ref{ee2}) for the electron neutrino survival
probability after vacuum oscillations into Eqs. (\ref{pro}) using
Eq. (\ref{pes}). And finally compare the results with the experimental
ratios $R(J)$  for each of
the relevant experiments ($J=H,K,G$).
{}From this procedure we can find the parameter region where oscillation
effects make the theoretical values of the survival solar
neutrino probability compatible  with the smaller than expected  solar neutrino
experimental flux.

In Fig.1.a we show the average probability as a function of $\delta m^2$
when $\sin^2\theta$ runs from $0.25$ (highest curve) to
$0.75$ (lowest curve). In Fig.1.b we show the same probability as
a function of $\sin^2\theta$ with
$\delta m^2$ running from $8\times 10^{-11}$ (highest curve) to
$3\times10^{-11}$ (lowest curve), using
Eqs.~(\ref{pro}). The allowed region in the $\delta m^2-\sin^2\theta$ plane
(at 95$\%$ c.l.) is displayed in  Fig.1.c for each of the three
experiences: Homestake (upper row),
Kamiokande (middle row) and Gallex (lower row).

In Fig. 2 we show the compatibility  region for the three experiences at
90$\%$ (Fig.2a) and $95\%$ (Fig.2b) of confidence level.

We have analysed the three generation neutrino oscillations in vacuum
as a possible solution to the solar neutrino problem. Fixing some of
the oscillating parameters (two mixing angles and one neutrino mass)
through the procedure described in Ref.~\cite{tau4}, we come to the
following conclusions. The mixing angle $\theta$ as well as the
squared mass difference $\delta m^2$ remain free parameters to be
used to fit the solar neutrino data and the theoretical neutrino
flux. The result obtained in Fig. 2 show that the values of these
parameters are $0.3\alt\sin^2\theta\alt0.7$ and
$3\times10^{-11}\,\mbox{eV}^2 \alt\delta
m^2\alt8\times 10^{-11}\,\mbox{eV}^2$. Interesting enough, such
values are of the same magnitude of those ones found in two
generation analysis~\cite{nh}. This can be understood remembering that
the large value of $m_3\sim165$ MeV implies very short wavelength $L$
(see Eq.(\ref{ee})) and consequently the effective oscillation occurs among
the two lightest generations. Nevertheless, the nonvanishing value of $\beta$
in Eq.(\ref{ee2}) guarantees that the three generation oscillation effects we
are analysing differ from the usual pure two generation oscillation
phenomenon. When we put $\beta=\gamma=0$ we obtain the usual two
generation results~\cite{nh}.

We have not addressed here the atmospheric neutrino problem because it is not
a well established experimental problem. Evenmore it is not clear that is
related to neutrino oscillations.

Finally, we would like to stress that although the numerical results we have
obtained in this work depend on the values of the angles  $\beta$ and $\gamma$
and of the mass $m_3$ we have used, our general approach remains valid even if
future experimental data imply in different values for these parameters.

\newpage
\acknowledgements

We thank Funda\c c\~ao de Amparo \`a Pesquisa do Estado de
S\~ao Paulo (FAPESP) for full financial support (O.L.G.P.) and
Con\-se\-lho Na\-cio\-nal de De\-sen\-vol\-vi\-men\-to Cien\-t\'\i
\-fi\-co e Tec\-no\-l\'o\-gi\-co (CNPq) for partial financial support
(M.M.G, V.P. and R.Z.F.).

\begin{figure}
\setepsfscale{0.7}
\begin{picture}(700,554)(0,0)
\end{picture}
\protect{\caption[]{Using Eqs.~\protect{\ref{pro}} we have plotted the
pro\-ba\-bi\-li\-ty transition as a
function of $\delta m^2$ (a), as a function of $\sin^2\theta$ (b) and
the contour plot at $95\,\%$ C.L. The Homestake data appears in the
first row, the Kamiokande ones in the middle row and the Gallex data
in the bottom row.}}
\label{fig1}
\end{figure}

\begin{figure}
\setepsfscale{0.7}
\begin{picture}(700,554)(0,0)
\end{picture}
\protect{\caption[]{Compatibility common region for the 3 experiences at
$90 \% \, C. L. $ (a)
and at $95 \%  \, C. L. $ (b).}}
\label{fig2}
\end{figure}

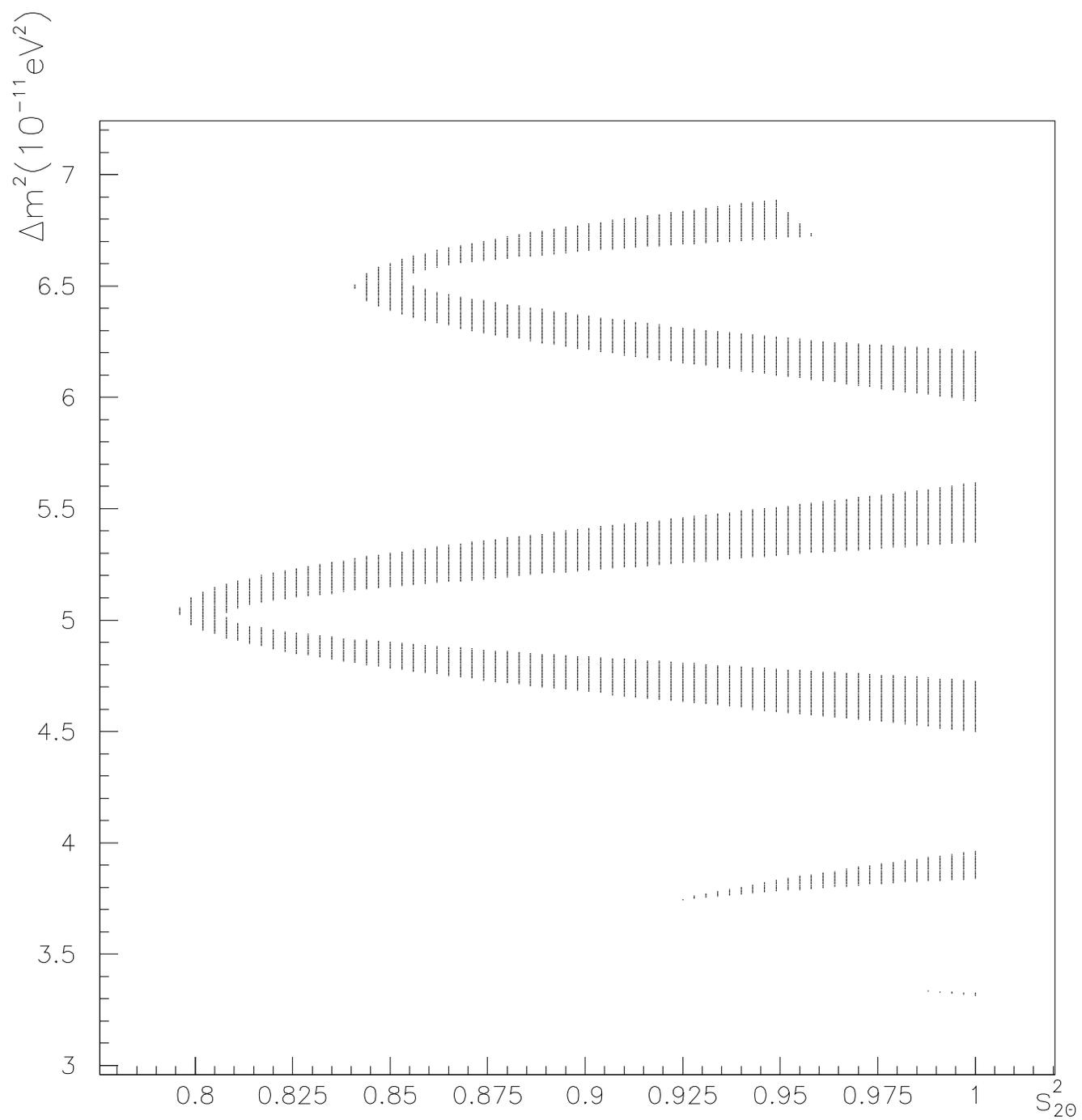
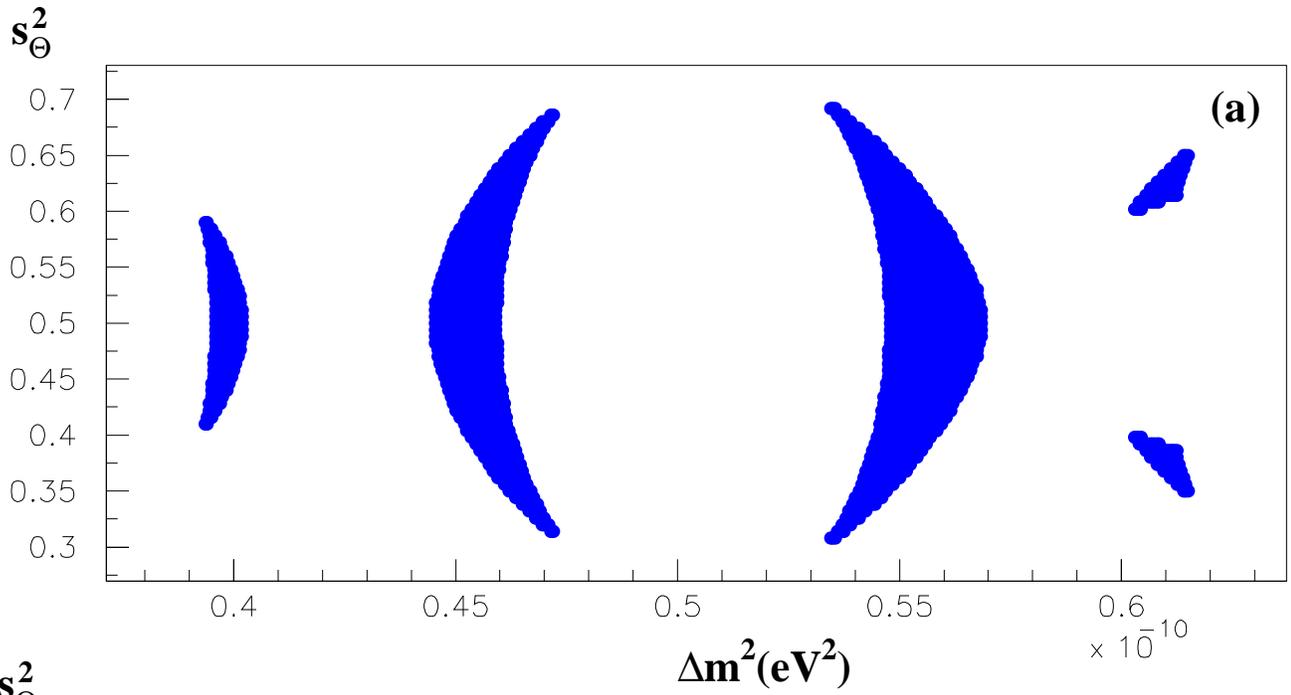
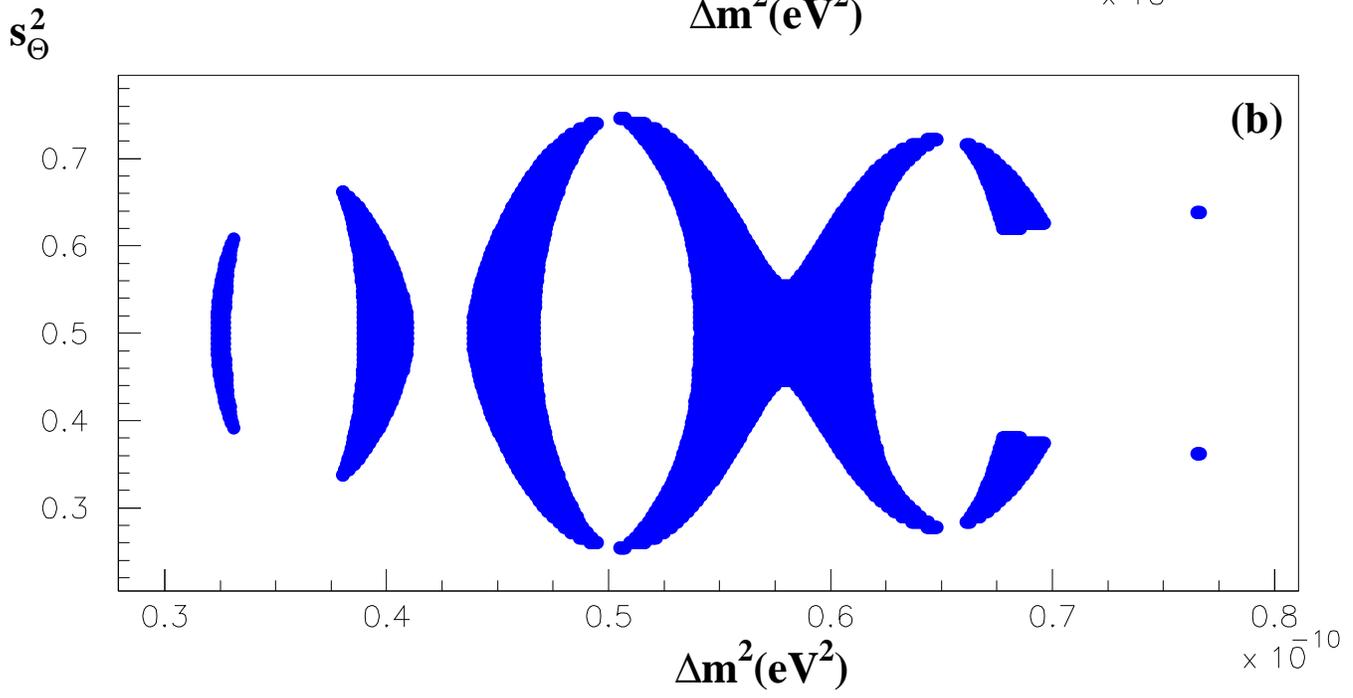
\end{document}